\documentclass[conference]{IEEEtran}
\IEEEoverridecommandlockouts
\usepackage{cite}
\usepackage{amsmath,amssymb,amsfonts}
\usepackage{multirow}
\usepackage{algorithmic}
\usepackage{graphicx}
\usepackage{textcomp}
\usepackage{xcolor}
\usepackage{flushend}
\usepackage{amsmath}
\usepackage{url}
\usepackage{balance}

\begin{document}

\title{Fast quantum circuit simulation using hardware accelerated general purpose libraries}

\author{\IEEEauthorblockN{Oumarou Oumarou}
\IEEEauthorblockA{\textit{Department of Informatics}\\
\textit{Clausthal University of Technology},  \\
38678 Clausthal-Zellerfeld, Germany}
\and
\IEEEauthorblockN{Alexandru Paler}
\IEEEauthorblockA{\textit{Department of Compute Science}\\
\textit{Transilvania University},\\
500036, Brașov, Romania}
\and
\IEEEauthorblockN{Robert Basmadjian}
\IEEEauthorblockA{\textit{Department of Informatics}\\
\textit{Clausthal University of Technology}, \\
38678 Clausthal-Zellerfeld, Germany}
}

\maketitle

\begin{abstract}
Quantum circuit simulators have a long tradition of exploiting massive hardware parallelism. Most of the times, parallelism has been supported by special purpose libraries tailored specifically for the quantum circuits. Quantum circuit simulators are integral part of quantum software stacks, which are mostly written in Python. Our focus has been on ease of use, implementation and maintainability within the Python ecosystem. We report the performance gains we obtained by using CuPy, a general purpose library (linear algebra) developed specifically for CUDA-based GPUs, to simulate quantum circuits. For supremacy circuits the speedup is around 2x, and for quantum multipliers almost 22x compared to state-of-the-art C++-based simulators.
\end{abstract}

\section{Introduction}

Quantum circuit simulation is receiving increased attention due to its applicability in the training of variational models of computation. For example, the parameters of a computational model can be adapted by simulating the circuit output values, computing the value of an error function and then updating the parameters such that the error function is minimised. Quantum circuit simulation is also used during the analysis of small-scale quantum circuit fault-tolerance or other structural properties. Furthermore, quantum circuit simulation has also become an important part of the quantum literacy efforts \cite{nita2020challenge}, because those tools (most of the times games) use simulators in the background.

Quantum circuit simulators are integral part of quantum software stacks. Most of the quantum software is built using Python, and our focus has been on ease of use, implementation and maintainability within the Python ecosystem. We were interested in building a high performance quantum circuit simulator using off the shelf, general purpose high performance libraries. The advantage of such libraries is their longer life compared to special purpose libraries and tools: due to their wide applicability there exists a larger community of developers and the risk of abrupt development interruption is minimised. Moreover, bugs and issues are discovered and repaired faster.

GPUs are no longer exotic hardware for general purpose computations: supercomputers and cloud infrastructures (e.g. Google Colab) use GPUs on a daily basis. Moreover, various types of GPU-based quantum circuit simulators have been developed (see the Related Work section). Nevertheless, due to the conflicting library versions and constraints in terms of programming languages (e.g. Python vs C++) or frameworks (e.g. Cirq\cite{cirq}, Qiskit\cite{qiskit}, PennyLane\cite{bergholm2018pennylane}) it is not always straightforward to use a GPU-based quantum circuit simulator.

\subsection{Background}

There are two challenges related to quantum circuit simulation: a) simulating quantum circuits as fast as possible; b) simulating the largest possible number of qubits. The latter is exponentially difficult, because the state of an $n$-qubit circuit is described by a $2^n$-sized complex vector called amplitudes, and each moment of a circuit's execution is represented by a $2^n \times 2^n$ complex matrix. For these reasons, arbitrary quantum circuits of at most 30 qubits can be simulated with consumer grade computers. For example, the simulation of a 42-qubit circuits would require 64TB RAM, which would easily occupy the memory of a supercomputer. Simulating all the amplitudes at once is specific for the so-called \emph{Schroedinger}-type quantum simulators. The simulation time of these simulators grows linearly with the number of gates in the circuit.

There are ways to avoid storing the entire state in the RAM \cite{pednault2019leveraging}, but there still seems to be a limit on the maximum number of qubits that a classical computer can hope to simulate in a universal arbitrary quantum computer. These are \emph{Feynman}-type simulators, e.g. \cite{villalonga2020establishing}, where only a few individual amplitudes are computed as the sum of the circuit paths that contribute to the value. The simulation time with these simulators grows exponentially with the size of the circuit. There are also hybrid techniques (state-vector simulators combined with path-sum simulators) which were used, for example, for the simulation of the supremacy circuits \cite{villalonga2020establishing}.

\subsection{Related Work}

This work focuses solely on a Schroedinger type simulator to be used on GPUs. Such simulators were proposed and implemented as early as \cite{lukacgpu, amariutei2011parallel}. However, their focus has not been on high performance computing and more capable simulators where the communication overhead was also analysed and modelled were implemented, for example, in \cite{jones2019quest, guerreschi2020intel}. Nevertheless, the fast advancement of GPU performance combined with the ever faster CPU-GPU interconnects meant that training variational quantum circuits on a laptop (RAM capacity permitting) \cite{bergholm2018pennylane} became feasible. One of the state of the art GPU-based simulators for such a task is Qulacs \cite{suzuki2020qulacs}.

We have witnessed how GPUs became standard building blocks of supercomputers (e.g. Summit, Tianhe-2A). The high performance computing community seems to have focused on the execution of Python code on the supercomputers. The goal is to write once execute on multiple platforms, such that NumPy code, for example, can be easily prototyped on a laptop and then executed on a huge machine. One example of libraries capable of being used in heterogeneous environments is NVidia Rapids \cite{hernandez2020performance}. 

NumPy is one of the most used Python libraries for performing array-based numerical computations. It runs on a single CPU core and is not generally parallelised. However, Legate \cite{bauer2019legate} was proposed as a programming system that transparently accelerates and distributes NumPy programs to machines of any scale and capability typically by changing a single module import statement. The same approach (replacing a single import statement) allows one to run NumPy code using CuPy\cite{nishino2017cupy} -- the CUDA GPU version of NumPy. Replacing libraries with a GPU version is not a Python-only approach: in \cite{avila2014gpu} the authors describe a Java-based CUDA quantum circuit simulator using this approach. Using drop-in off the shelf library replacements has the advantage that it simplifies the maintainability and and stability of the code base.

\section{Methods}
\label{sec:methods}

In this paper we investigate the feasibility of using CuPy\cite{nishino2017cupy} for writing a fast, versatile GPU-based quantum circuit simulator. CuPy is the NumPy equivalent library that supports CUDA enabled GPUs, and considering \cite{hernandez2020performance} it has the potential to be used in high performance computing environments with more than a single GPU.

We start from the assumptions that: a) in its simplest incarnation a quantum circuit simulator is a matrix vector multiplication software, b) the supporting libraries should be as easy as possible to install on consumer and specialised hardware; c) the performance is important for large qubit numbers when memory and communication overhead seem to be the bottleneck (ie. \cite{willsch2021gpu} for a very recent discussion of GPU simulators and the associated communication overhead). Our goal is to analyse the performance of simulating quantum circuits with GPUs. While this approach is already standard in the literature, we are interested in using off the shelf libraries for building the simulator as well as for benchmarking the obtained performance.

We develop our own quantum circuit simulator and include it in QUANTIFY \cite{quantify}. Our simulator uses the Cirq state vector simulator as a starting point and we exploit the speed of GPUs by replacing NumPy calls with CuPy ones. This is possible and straightforward due to the API of the latter. For our code, we had to fix some NumPy type errors in the Cirq/linalg packaged, but the latest versions of Cirq already include these repairs.

It should be noted that our methodology to construct the CuPy-based simulator guarantees that the simulated circuit outputs are valid as long as CuPy is performing correct. Our simulator has all the benefits of the standard Cirq simulator, and is fast too. In the following \emph{we will refer to the Cirq NumPy-based simulator as NumPy, and we will call our simulator CuPy}.

We benchmarked the performance of CuPy using two types of circuits: supremacy and arithmetic. We use QUANTIFY \cite{quantify} to generate quantum arithmetic circuits (e.g. multipliers). QUANTIFY is a Cirq-based open-sourced framework capable of (1) compiling state-of-the-art quantum circuits (it has a large library of implemented circuits), (2) decomposing the circuits into their constituent different types of gate-levels, (3) optimising the decomposed circuits based on some heuristics-based rules, and (4) verifying either the decomposed or the optimised circuits against their mathematical models using state-of-the-art adopted metrics (e.g. T-count, T-depth, CNOT-count, etc.). For the experiments related to the supremacy circuits, we use the Google-based Cirq framework \cite{cirq}.

The scaling of the analysed circuit widths is polynomial (a multiple of the problem size), but we were interested in a finer grained analysis one qubit-by-qubit basis. We want to investigate performance scaling for any number of qubits. For this reason, whenever we wanted to simulate a circuit of a non-standard width (e.g. supremacy circuit with 19 qubits), we generate the next largest circuit (e.g. 20 qubits) and randomly removed qubits (and the corresponding gates) until achieving the required width.

\section{Performance Evaluation}

We present the performance (the overall execution time of the experiment expressed in seconds) obtained by executing different experiments related to two types of quantum circuits namely supremacy \cite{supremacy} and multipliers \cite{thapliyal}. We are interested in comparing the performance of different simulators and consider NumPy state-vector simulator, our CuPy simulator, QSim \cite{qsim} (called and executed from Cirq) and Qulacs (called and executed from Cirq). For the latter two simulators we analyse their performance when called through the Python-bindings. The simulators could expose a higher speed when used from the command line but, for example, when training variational models, the simulators are called through the Python interfaces. We are interested in the simulation speeds achieved in practical setup.

In order to keep the comparison fair between the simulators we use the same Python quantum software, namely Cirq. We simulate each circuit 10 times and average the execution times. The simulation environment are Google Colab cloud machines with Tesla T4 GPUs (randomly assigned to our machine) on PCI 16x with 16GB RAM, CUDA Version 11.2, while the machine had 12 GB RAM and a Xeon(R) CPU @ 2.20GHz. Initially, we simulated the multipliers with randomly selected binary inputs, but then decided to use the same supremacy-circuit-like approach. Both supremacy and arithmetic circuits were simulated with equal superposition inputs. For the first this is the standard approach, while for the latter we wanted to make sure that CuPy is not optimising the CPU-GPU transfers due to state vector sparsities. 

In a nutshell, the CuPy simulator performs best for larger circuits. Our hypothesis for the low performance for low qubit counts (21 qubits for supremacy in Figure~\ref{fig:supremacy}, and 15 qubits for multiplication in Figure~\ref{fig:muliplier}) is that the CPU-GPU communication overhead is higher than the computation speedup.

\subsection{Multiplication circuits}

For the multiplication circuits, we considered the ones presented by \cite{thapliyal}. These have a total width of $4n+1$ such that $2<n<8$. The parameter $n$ represents the size of the integer inputs to be multiplied. The depth of the circuit is $\mathcal{O}(n^2)$.

If we consider only the input integers' size $n$ as the free parameter for the horizontal axis, then we would get relatively sparse data. To mitigate this problem, we use the method described in the previous section. For $2<n<8$ we increasingly and randomly take out qubits from the multiplier circuits of width $4n+1$ and $4(n+1)$ so that the circuit size sweeps over the quantities between both widths. Consequently, we obtain no gaps or jumps between both quantities but rather a continuous circuit width which takes all the values in between. For our upper limit which is $4\times7+1$, we were unable to simulate the multiplier due to memory constraints. We simulate the circuit for the values between $25$ and $28$ qubits by removing those qubits from the circuits in the same fashion as described above.

\begin{figure}[t!]
    \centering
    \includegraphics[width=0.48\textwidth]{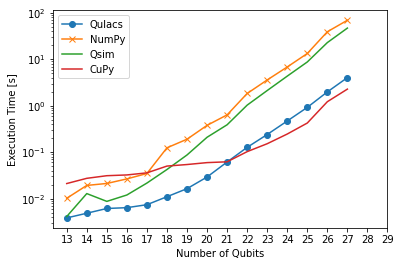}
    \caption{The speed of NumPy, CuPy, QSim and Qulacs for \emph{multiplication circuits} having a width between 4 and 28 qubits.}
    \label{fig:muliplier}
\end{figure}

Figure \ref{fig:muliplier} illustrates the results of the experiments for the use case of the multiplier circuit, by considering 4 different simulation configurations and the case of 13 till 28 qubits (the limit before obtaining out of memory error). Note that for the case of Qulacs simulation configuration, the upper limit was 25 qubit after which we obtained out of memory errors. It is important to mention that the experiments were repeated 10 times and their averages were calculated. Among the four simulation configurations of NumPy, CuPy, QSim and Qulacs, for the case of 13 till 17 qubits all the four simulation configurations have similar execution times with Qulacs having slightly the edge over the others. After 17 qubits, it becomes evident that our choice of CuPy owns better performance (e.g. smaller execution times) than the other three configurations especially by increasing the number of qubits. More precisely, we noticed that CuPy is in factor of 2 and 10 faster than Qulacs and NumPy as well as QSim respectively.

\subsection{Supremacy circuits}

The supremacy circuits for benchmarking were used to assess for the supremacy of the quantum computation model over the traditional counterpart. We worked with square-shaped circuits of size $n \times n$ qubits. The maximum number of qubits we succeeded at simulating with particular setting (square circuits) was $5 \times 5= 25 $ qubits, because $6 \times65= 36$ qubits cannot be stored in a usual computer RAM. Using our type of supremacy circuits, we achieved a limit of $7\times 4 = 28$ qubits. Intermediate circuit sizes were generated using the method described in Section~\ref{sec:methods}.

\begin{figure}[t!]
    \centering
    \includegraphics[width=0.48\textwidth]{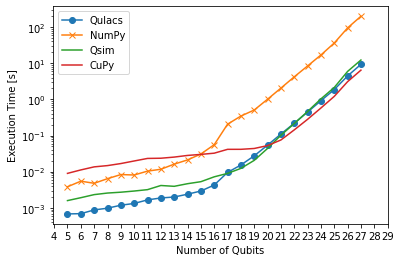}
    \caption{The speed of NumPy, CuPy, QSim and Qulacs for \emph{supremacy circuits} having a width between 13 and 28 qubits.}
    \label{fig:supremacy}
\end{figure}

Figure \ref{fig:supremacy} presents the obtained results of the carried out experiments for the use case of the supremacy circuit, by considering 4 different simulation configurations and the case of 5 to 28 qubits. Note that the supremacy circuit has a much less memory requirements than the multiplier circuit. Hence, for the case of Qulacs simulation configuration, we did not face the same problem as we had for the multiplier circuit. It is important to mention that the experiments were repeated 10 times and their averages were calculated and demonstrated in the figure. Among the four simulation configurations of NumPy, CuPy, QSim and Qulacs, for the case between 4 and 15 qubits all the four simulation configurations have similar execution times with Qulacs having slightly the edge over the others. For more than 15 qubits, it becomes evident that our choice of CuPy has slower execution times than the other three configurations, whereas NumPy has the worst performance.

\begin{figure}[t!]
    \centering
    \includegraphics[width=0.48\textwidth]{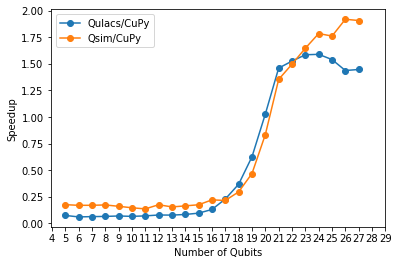}
    \caption{The speedup compared to Qulacs and Qsim achieved with CuPy for \emph{supremacy circuits} having a width between 5 and 28 qubits.}
    \label{fig:ratios}
\end{figure}

\begin{figure}[t!]
    \centering
    \includegraphics[width=0.48\textwidth]{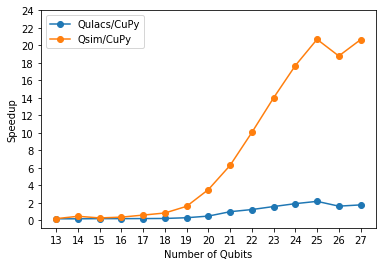}
    \caption{The speedup compared to Qulacs and Qsim achieved with CuPy for \emph{multiplication circuits} having a width between 13 and 28 qubits.}
    \label{fig:ratiom}
\end{figure}

\section{Discussion and Conclusion}

We presented how to massively improve the performance of a state vector quantum circuit simulator by using a drop-in replacement library for GPU acceleration of linear algebra mathematics, namely CuPy. Our approach is motivated by the fact that we are interested in developing a stable and maintainable quantum software stack where the simulators are both fast, versatile and capable to be easily operated on heterogeneous hardware. Source code is available at \url{http://www.github.com/quantumresource/quantify}.

The really large speedups achieved using CuPy (Figures~\ref{fig:ratios} and \ref{fig:ratiom})  was achieved without explicitly calling cuSPARSE which is a part of the CuPy library. Nevertheless, the NumPy-based simulator from Cirq exploits matrix sparsity. It could be possible that cuSPARSE will further improve the speedup, but this is not guaranteed.

We did not perform any particular optimisation to increase the GPU throughput, and we did not explicitly benchmark the memory consumption between the simulators. Such tasks are future work. However, while benchmarking Qulacs we noticed increasing memory footprints.

We were surprised by the performance gains achieved for supremacy circuits (approx. 2x compared to Qsim), but even more impressive were the speedups achieved for the simulation of arithmetic circuits. Our approach has the potential to be easily scaled to high performance machines because the underlying libraries were designed and implemented accordingly. Future work will focus on simulations with multiple GPUs.

\bibliographystyle{plain}
\bibliography{__main}

\end{document}